\documentclass[a4paper]{jpconf}
\usepackage{iopams}
\usepackage{graphicx}
\usepackage{cite}	
\usepackage{floatflt}

\begin{document}
\title{Magneto-thermal properties
of the spin-$s$ Heisenberg antiferromagnet on the cuboctahedron}

\author{A Honecker$^1$ and M E Zhitomirsky$^2$}

\address{$^1$ Institut f\"ur Theoretische Physik,
 Universit\"at G\"ottingen, 37077 G\"ottingen, Germany}

\address{$^2$ Commissariat \`a l'Energie Atomique,
 DSM/INAC/SPSMS, 38054 Grenoble, France}

\ead{ahoneck@uni-goettingen.de}

\begin{abstract}
We use the example of the cuboctahedron, a highly frustrated molecule with 
12 sites, to study the approach to the classical limit. We compute 
magnetic susceptibility, specific heat, and magnetic cooling rate at high 
magnetic fields and low temperatures for different spin quantum numbers 
$s$. Remarkably big deviations of these quantities from their classical 
counterparts are observed even for values of $s$ which are usually 
considered to be almost classical.
\end{abstract}

\section{Introduction}

\begin{floatingfigure}[p]{0.31\columnwidth}
\begin{minipage}[b]{0.27\columnwidth}
\begin{center}
\includegraphics[width=0.8\columnwidth]{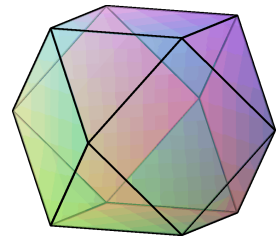}
\end{center}
\caption{\label{Fig:cuboc}
The cuboctahedron: a highly frustrated magnetic molecule with
12 spin sites.}
\end{minipage}
\end{floatingfigure}

\noindent
Efficient low-temperature cooling by adiabatic demagnetization is a 
potential application of highly frustrated antiferromagnets 
\cite{Z03,SPSGBPBZ05,ZH04,HW06,STS07}. However, quantitative theoretical 
results for the magneto-thermal properties have so far been obtained only 
in the classical case \cite{Z03,SPSGBPBZ05} and the extreme quantum case 
of spin $s=1/2$ \cite{ZH04,HW06,STS07}, with the exception of a study of 
the magnetocaloric effect in a (non-frustrated) ferrimagnetic 
spin-$(5/2,1)$ chain \cite{BBO07}. Note that numerical computations are 
already quite demanding for highly frustrated spin-1/2 antiferromagnets in 
low dimensions and become quickly unfeasible for $s > 1/2$ or in higher 
dimensions.

The cuboctahedron (see Fig.~\ref{Fig:cuboc}) is a highly frustrated 
molecule which can be considered as a variant of the kagome lattice with 
12 sites. At high magnetic fields, it supports localized magnon 
excitations \cite{SSRS01,SHSRSS02} which can be understood to be 
responsible for the enhanced magnetocaloric effect in these frustrated 
quantum spin systems. Several works have used the cuboctahedron as a model 
system for studying magneto-thermal properties \cite{SRS05,SRS07,RLM08}. 
Here we continue along these lines and use the cuboctahedron to 
investigate the behavior as a function of the spin quantum number $s$.

\section{Model, observables and method}

We consider the Heisenberg antiferromagnet
in an external magnetic field $h$
\begin{equation}
H = J \sum_{\langle i, j \rangle} \vec{S}_i \cdot \vec{S}_j - h \sum_i S_i^z \, ,
\label{Hop}
\end{equation}
where the $\vec{S}_i$ are {\it quantum} spin-$s$ operators and
$J > 0$ is the exchange constant between nearest neighbor
sites $\langle i, j \rangle$.
The {\it classical} Heisenberg model is obtained
from (\ref{Hop}) in the limit $s \to \infty$ at fixed
$J_{\rm cl} = s^2\,J$, $h_{\rm cl}=s\,h$.
In this classical limit the operators $\vec{S}_i$
can be considered as unit vectors.
Accordingly, a thermodynamic quantity ${\cal Q}_s$ for quantum spin $s$
approaches the classical limit as
\begin{equation}
s^{\alpha_{\cal Q}}\,
{\cal Q}_s\left(\frac{T}{s^2\, J}, \frac{h}{s\,J}\right)
\ \mathop{\longrightarrow}_{s \to \infty}\
{\cal Q}_{\rm cl.} \left(\frac{T}{J_{\rm cl}}, \frac{h_{\rm cl}}{J_{\rm cl}}\right) \, .
\label{obsCl}
\end{equation}
Using their definitions, it is easy to show that $\alpha_\chi = 0 = \alpha_C$,
$\alpha_M = -1$, and $\alpha_{\Gamma_h}=1$, where $\chi$, $C$, and $M$
are the magnetic susceptibility, specific heat, and magnetization, respectively.
Due to the discrete level spacing, quantum effects will be important
below a crossover scale $E_{\rm q} \propto s\,J$, {\it i.e.}, for
$T \lesssim E_{\rm q}$. At temperatures $T \gtrsim E_{\rm q}$,
${\cal Q}_{\rm cl.}$ is expected to yield an approximation
to ${\cal Q}_s$ after rescaling according to Eq.~(\ref{obsCl}).
The absolute value of the crossover scale $E_{\rm q}$ is
unknown, but one may hope that it is small in highly frustrated
magnets because of the generic reduction of energy scales.

\begin{figure}[p]
\includegraphics[width=0.47\columnwidth]{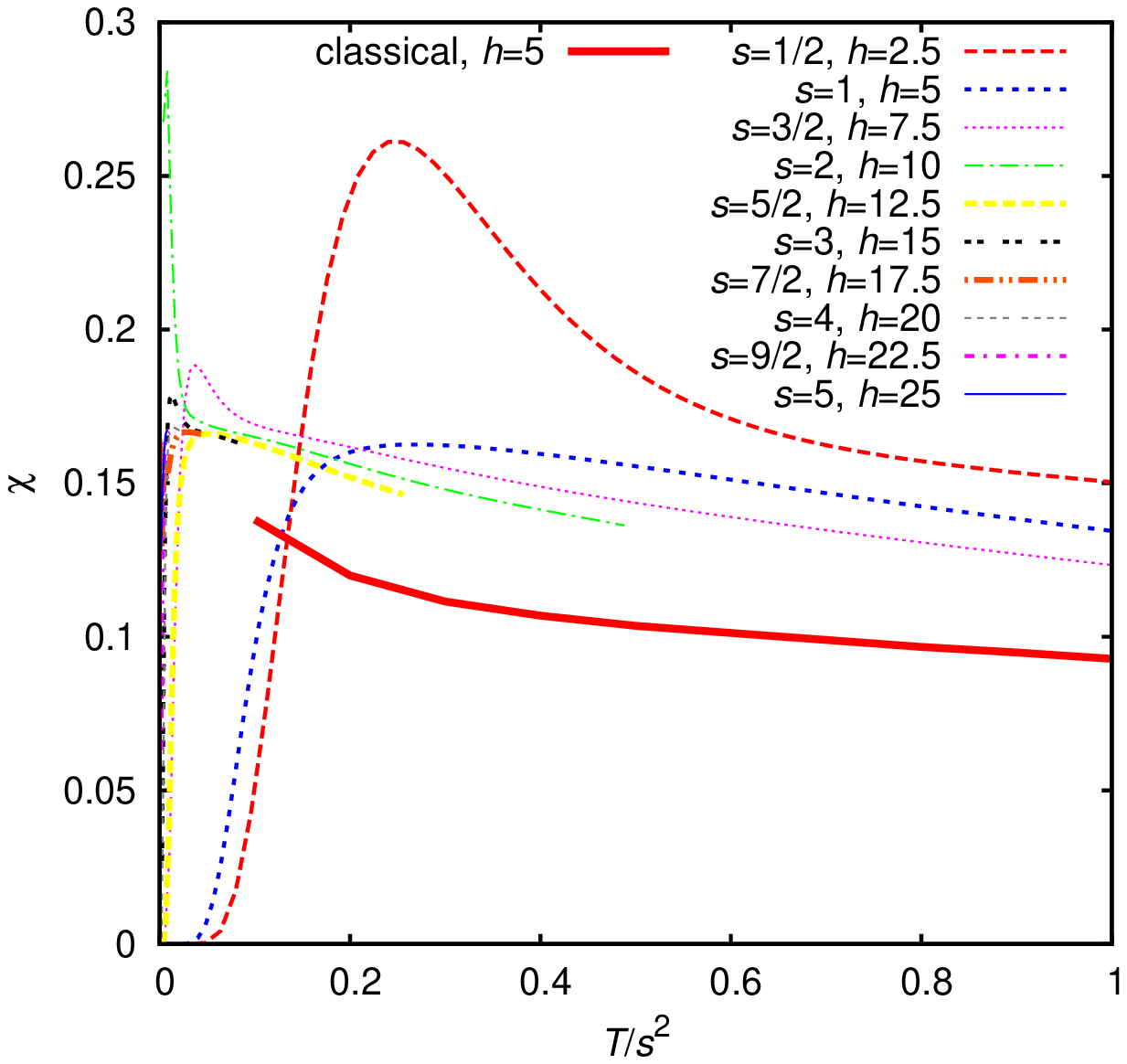}\hfill
\includegraphics[width=0.47\columnwidth]{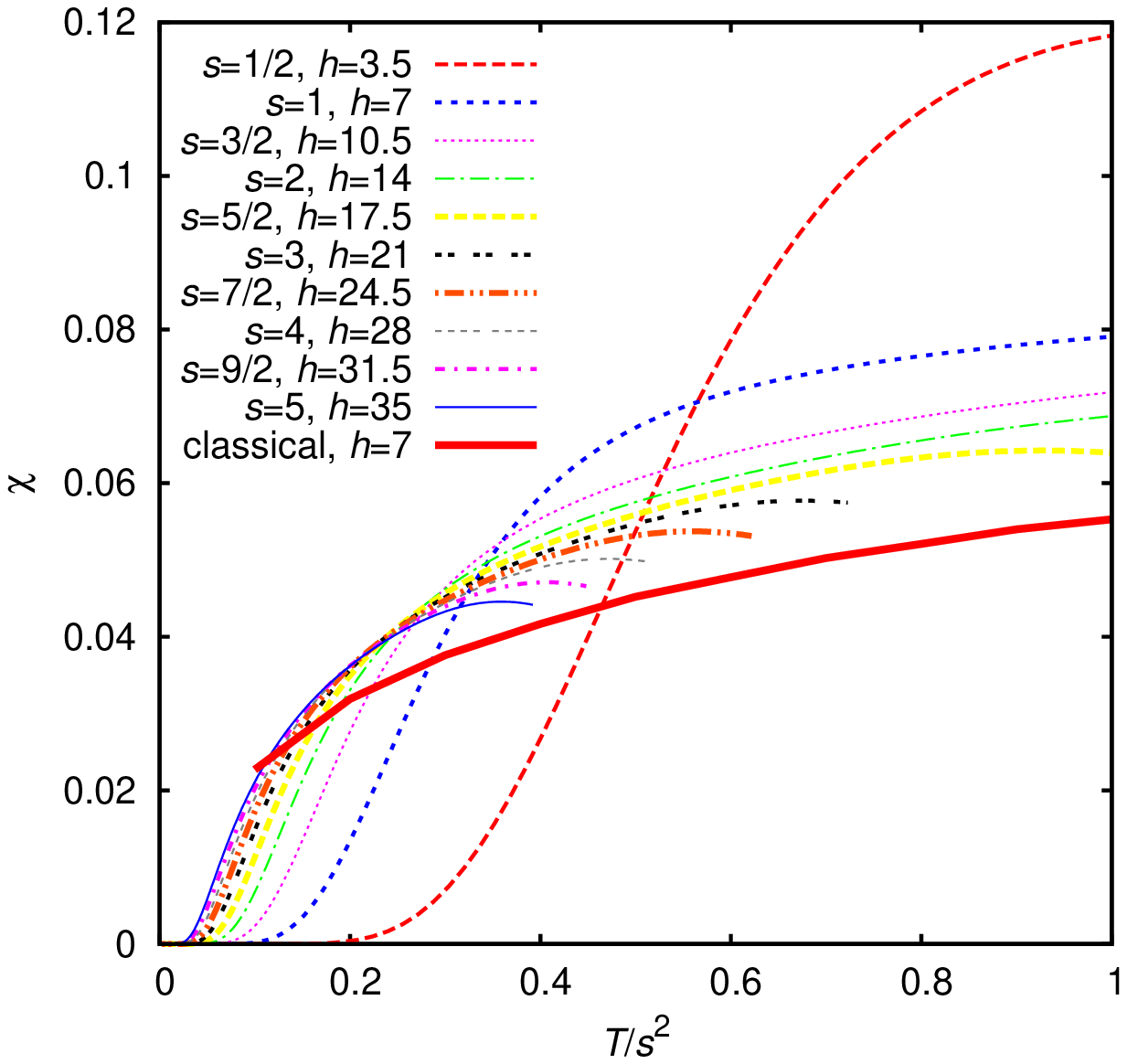}\\
\includegraphics[width=0.47\columnwidth]{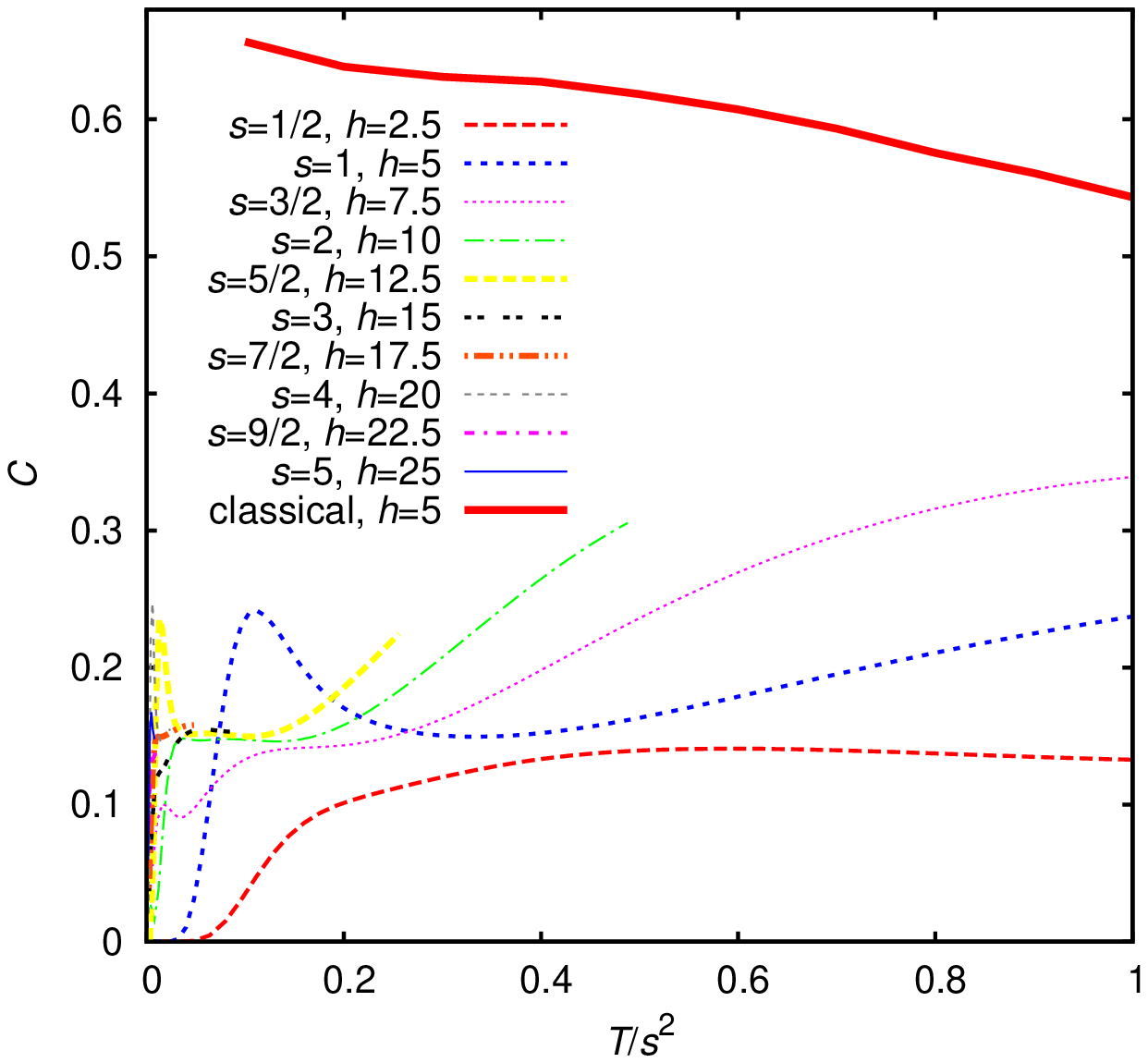}\hfill
\includegraphics[width=0.47\columnwidth]{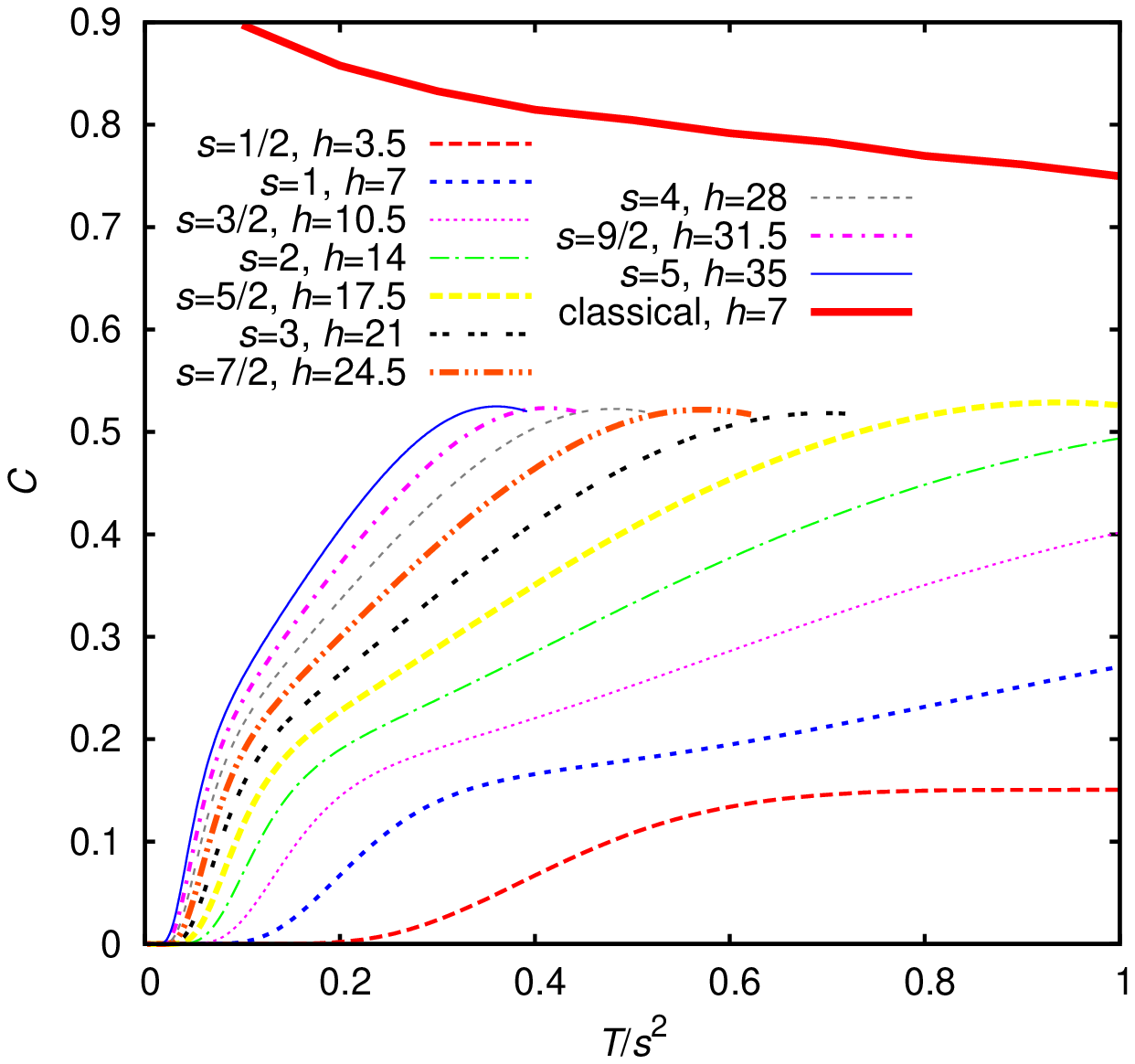}\\
\includegraphics[width=0.47\columnwidth]{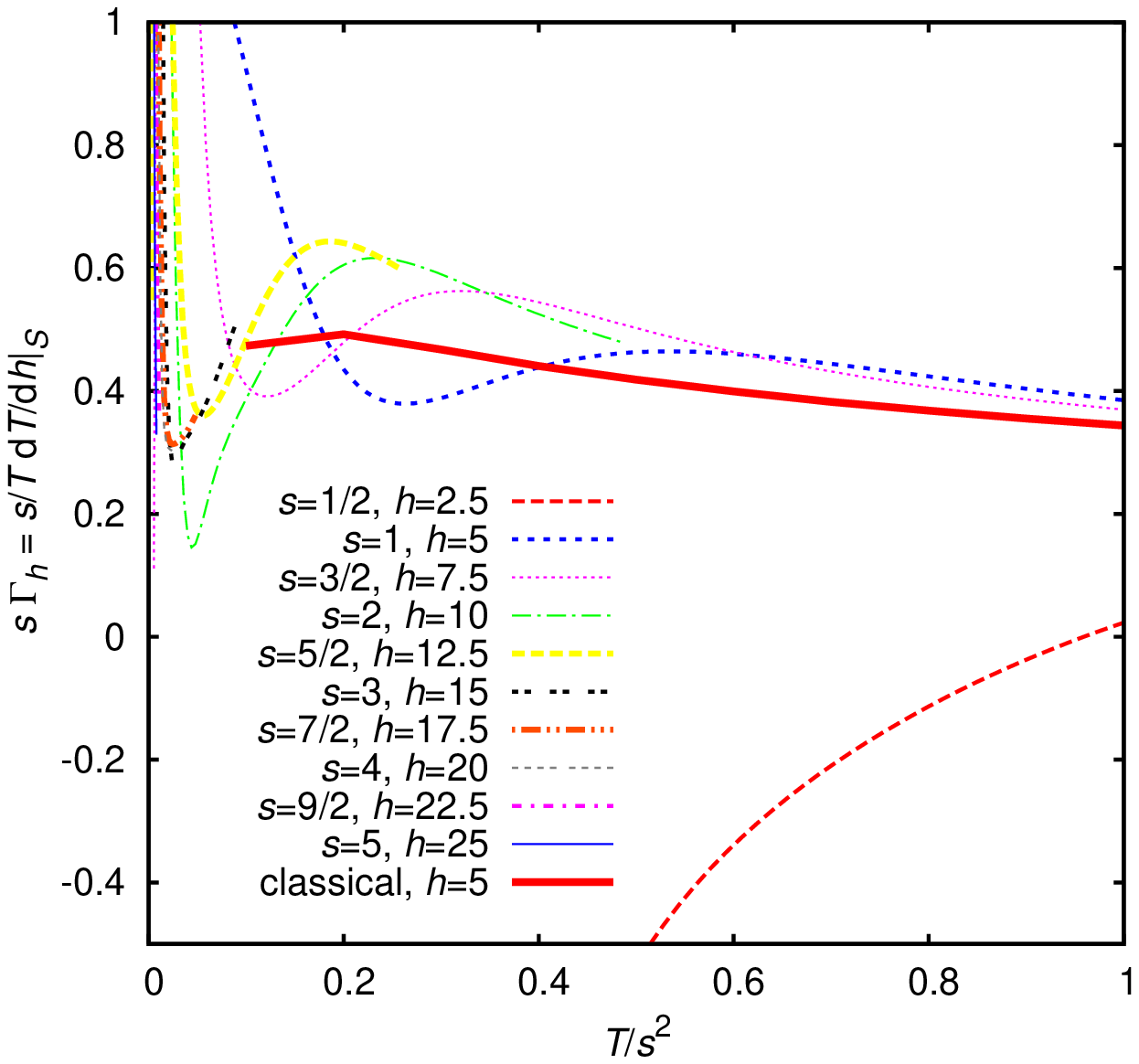}\hfill
\includegraphics[width=0.47\columnwidth]{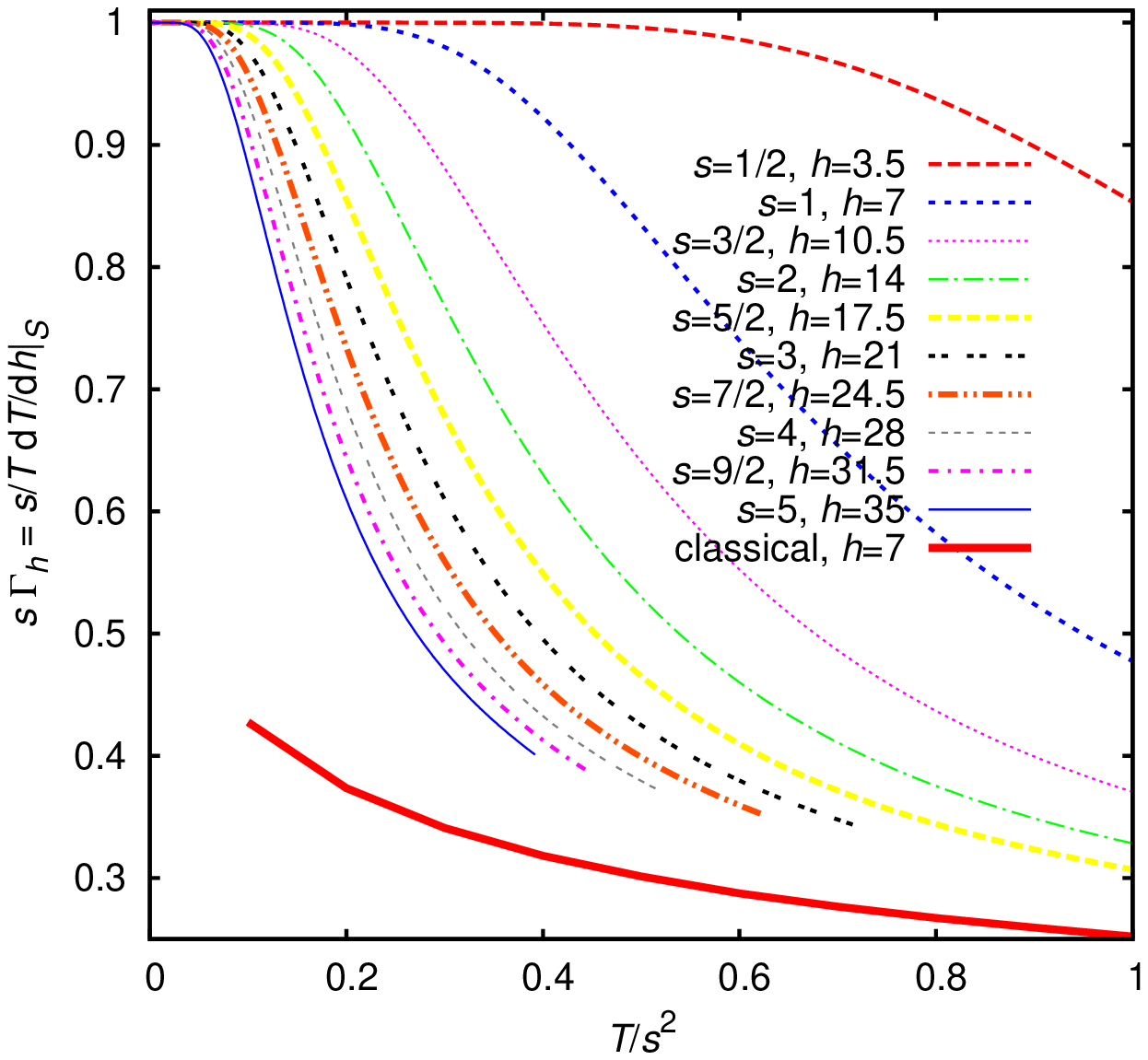}
\caption{\label{Fig:H57}
Magneto-thermal properties slightly below (left column)
and slightly above (right column) the saturation field:
magnetic susceptibility (top),
specific heat (middle), and
cooling rate $\Gamma_h$ (bottom).
All quantities are normalized to the number of spins (12).
}
\end{figure}

\begin{figure}[t]
\includegraphics[width=0.47\columnwidth]{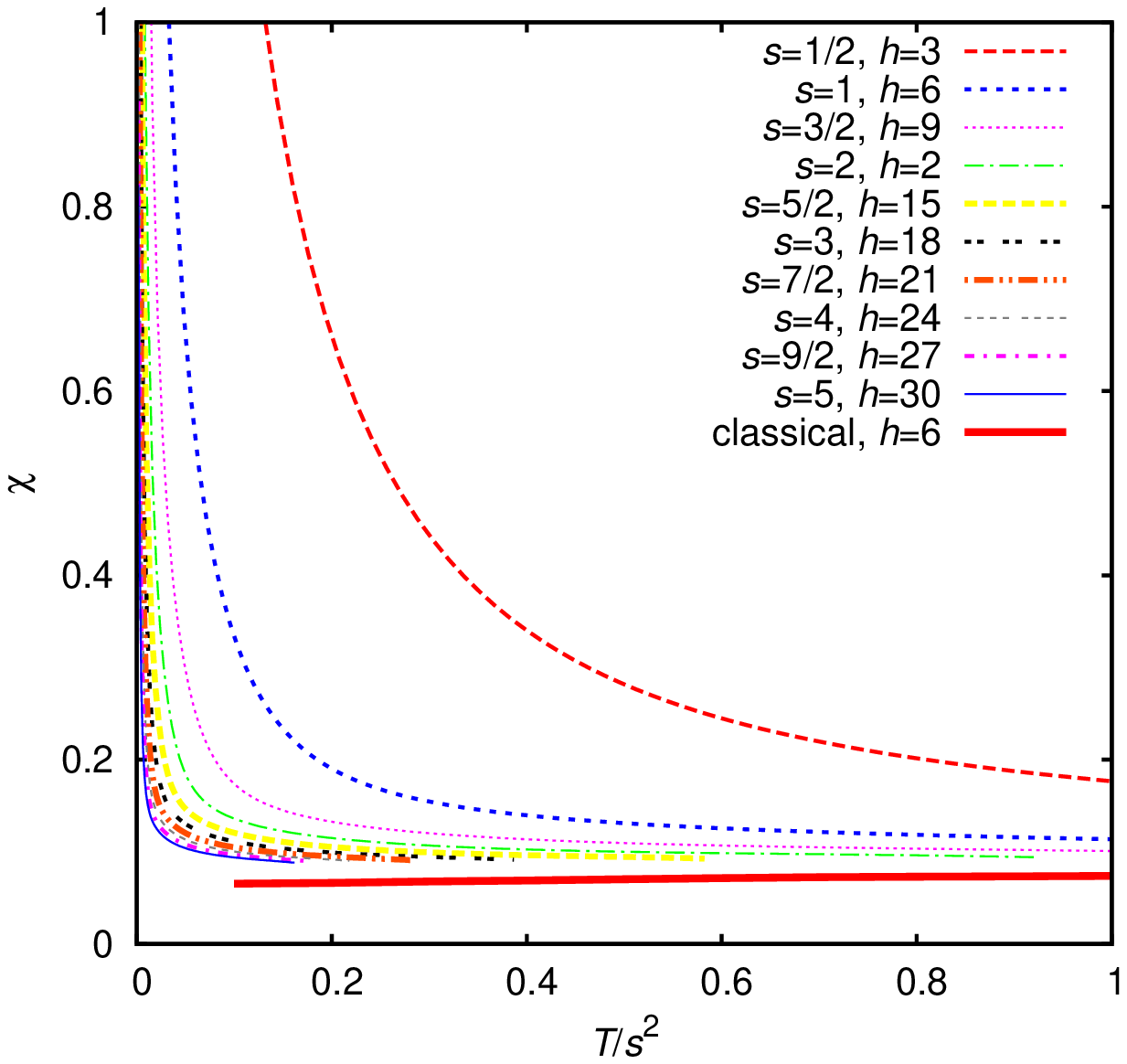}\hfill
\includegraphics[width=0.47\columnwidth]{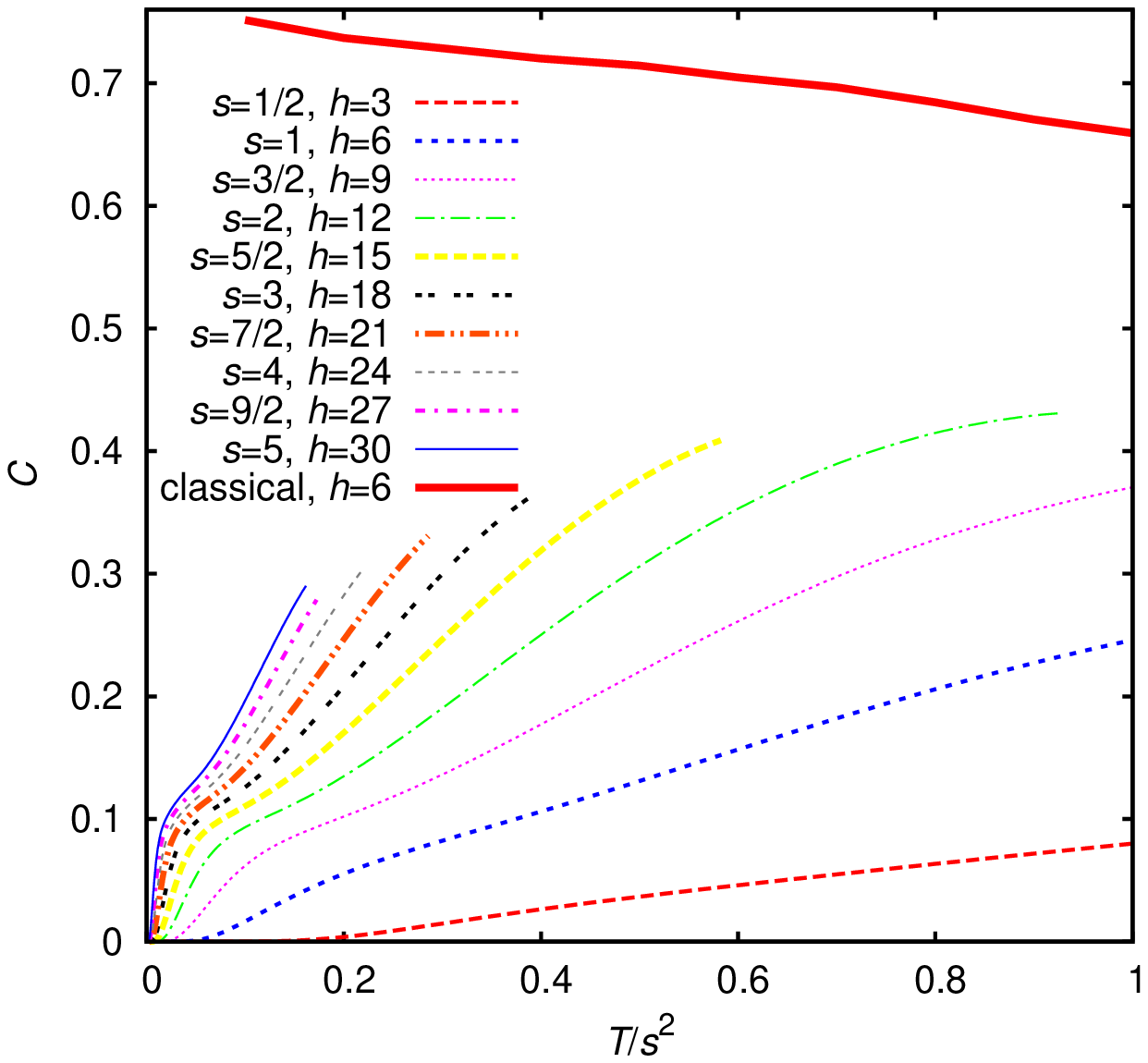}\\
\begin{minipage}[b]{0.47\columnwidth}
\caption{\label{Fig:H6r}
Magneto-thermal properties exactly at the saturation field:
magnetic susceptibility (top, left),
specific heat (top, right), and
cooling rate $\Gamma_h$ (right).
All quantities are normalized to the number of spins (12).
}
\vspace*{2.8cm}
\end{minipage}
\hfill
\begin{minipage}[b]{0.47\columnwidth}
\includegraphics[width=\columnwidth]{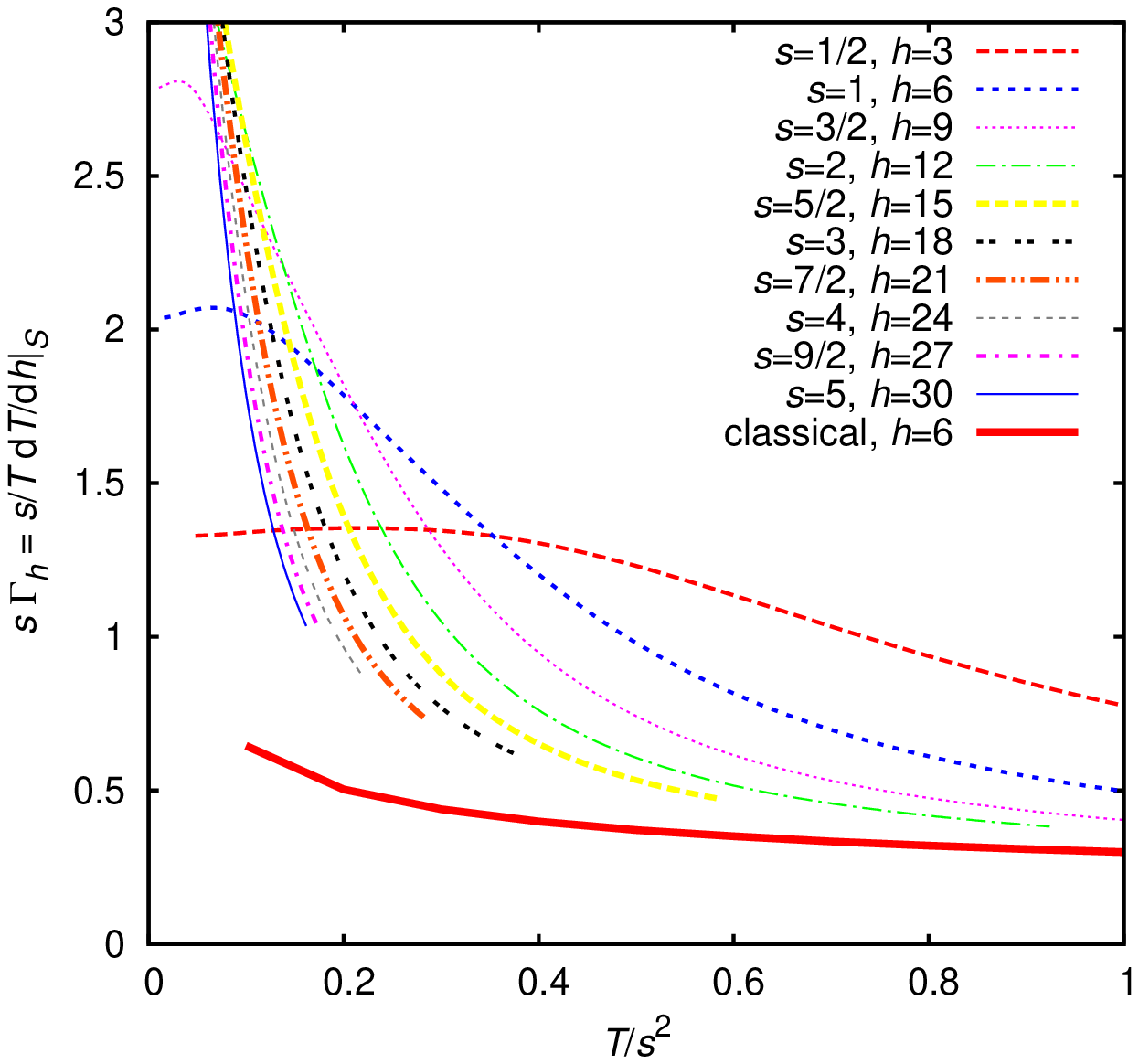}
\end{minipage}
\vspace*{-5mm}
\end{figure}

We will be particularly interested in  the adiabatic cooling rate
at fixed entropy $S$
\begin{equation}
\Gamma_h = \frac{1}{T} \, \left.\frac{{\rm d}T}{{\rm d}h}\right|_S
= - \frac{1}{C} \, \left.\frac{{\rm d}M}{{\rm d}T}\right|_h \, .
\label{adCool}
\end{equation}

Thermodynamic quantities can be computed for the cuboctahedron by
full diagonalization for $s=1/2$ and $1$. For $s \ge 3/2$ we have to
resort to a truncation of the spectra to low energies
and
one should be careful about truncation artifacts.
Therefore we consider only temperatures $T \lesssim E_{\rm c}/10$
if the spectrum is complete until an excitation energy $E_{\rm c}$
even if we have partial results also for higher energies.
For the classical case we use
a combination of the Metropolis Monte Carlo algorithm with
over-relaxation steps.

\section{Results}

Now we present some results for the low-temperature region $T \le s^2\,J$,
choosing $J=1=J_{\rm cl}$ for simplicity.
The behavior close to the saturation field $h = 6\,s$ is shown 
Figs.~\ref{Fig:H57} and \ref{Fig:H6r}. We can see a trend of convergence 
towards the classical limit, in particular in the cooling rate $\Gamma_H$. 
However, quantitative corrections remain large for most cases, even for 
spin quantum numbers which are as big as $s=5$. The least systematic 
behavior at low temperatures is seen in Fig.~\ref{Fig:H57} for $h = 5\,s$ 
where the system becomes gapless in the limit $s \to \infty$. A gap opens 
above the saturation field which leads to the thermal activation of $\chi$ 
and $C$ observed in Fig.~\ref{Fig:H57} for $h = 7\,s$. Exactly at the 
saturation field, a total of 9 states with different magnetic quantum 
numbers becomes degenerate at $T=0$ \cite{SSRS01,SRS07,RLM08}. This leads 
to the $1/T$ singularity in $\chi$ and the big values of the cooling rate 
$\Gamma_h$ which can be observed in Fig.~\ref{Fig:H6r}. For $s=1/2$ and 
$h=3$ we reproduce previous results for the specific heat $C$ 
\cite{SRS05}. Both the thermal activation at $h = 7\,s$ as well as the 
$1/T$ singularity in $\chi$
at $h=6\,s$
are quantum effects which disappear in the classical limit.

\section{Discussion}

Remarkably big deviations from the classical behavior are observed for 
magneto-thermal properties of the cuboctahedron with $s \le 5$. This 
should not come as a complete surprise since a slow convergence towards 
the classical limit has already been observed for the zero-field magnetic 
susceptibility of spin chains \cite{KGWB98} and certain magnetic molecules 
\cite{ELS06}.

It should be noted that the cuboctahedron may not be the most 
representative case. Indeed, it has been pointed out \cite{RLM08} that 
quantum effects are more pronounced in the cuboctahedron than in the 
icosidodecahedron which can be considered as a 30-site variant of the 
kagome lattice. It would therefore be desirable to study the approach to 
the classical limit for non-frustrated lattices where Quantum Monte Carlo 
simulations can be used to compute magneto-thermal properties on bigger 
systems. In the meantime, one should be careful about the applicability of 
purely classical models for a quantitative description of highly 
frustrated antiferromagnets in the low-temperature region
$T \lesssim s^2\,J$.

\ack
We would like to thank Johannes Richter for useful discussions and
comments on the manuscript.
A~H acknowledges financial support by the German Science Foundation
(DFG) through SFB602 and a Heisenberg fellowship (Project HO~2325/4-1).
We are grateful for allocation of CPU time on high-performance computing
facilities at the TU Braunschweig and HLRN Hannover.

\end{document}